\documentclass[a4paper]{article}

\usepackage[T1]{fontenc}
\usepackage[utf8]{inputenc}
\usepackage{authblk}
\usepackage[pdfstartview={FitH}]{hyperref}
\usepackage{amsmath,amssymb}
\usepackage{mathtools}
\usepackage{commath}
\usepackage{gensymb}
\usepackage{cancel}
\usepackage{booktabs}
\usepackage{graphicx}
\usepackage[font=small,labelfont=bf]{caption}

\newcommand{\overbar}[1]{\mkern 1.5mu\overline{\mkern-1.5mu#1\mkern-1.5mu}\mkern 1.5mu}

\begin{document}

\title{\textbf{Sensitivity of the JEM-EUSO telescope to gravity effects in neutrino-induced air showers}}
\author{Stefan Mladenov$^{1}$\thanks{Corresponding author: \texttt{smladenov@phys.uni-sofia.bg}} , Galina Vankova$^{1}$, Roumen Tsenov$^{1}$, Mario Bertaina$^{2}$, Andrea Santangelo$^{3}$}
\affil{$^{1}$\textit{Sofia University St. Kliment Ohridski, Sofia, Bulgaria}}
\affil{$^{2}$\textit{University of Turin and INFN Turin, Torino, Italy}}
\affil{$^{3}$\textit{Eberhard Karls University, Tübingen, Germany}}
\affil{for the JEM-EUSO Collaboration}
\date{}
\maketitle

\begin{abstract}
We examine the JEM-EUSO sensitivity to gravity effects in the context of Randall-Sundrum (RS) model with a single extra dimension and small curvature of the metric. Exchanges of reggeized Kaluza-Klein gravitons in the $t$-channel contribute to the inelastic cross-section for scattering of ultra-high-energy neutrinos off nucleons. Such effects can be detected in deeply penetrating quasi-horizontal air showers induced by interactions of cosmic neutrinos with atmospheric nucleons. For this reason, we calculate the expected number of quasi-horizontal air showers at the JEM-EUSO observatory as a function of two parameters of the RS model.
\end{abstract}

\section{Introduction}

One of the open questions in the modern particle physics is the so-called hierarchy problem, namely inexplicably large ratio of the gravity scale ($10^{19}$ GeV) to the electroweak scale ($10^2$ GeV). A whole model framework, the Arkani-Hamed--Dimopoulos--Dvali (ADD) model \cite{ArkaniHamed:1998rs,Antoniadis:1998ig,ArkaniHamed:1998nn}, was developed to explain the weakness of gravity relating to other forces by introducing additional spatial dimensions which are large compared to the Planck scale. This approach reduces the problem to fine tuning of few new parameters, i.e. the size of the additional dimensions.

The most minimalistic and elegant model which offers a solution to the problem is the Randall-Sundrum (RS) model \cite{Randall:1999ee} with a single extra dimension and warped background metric
\begin{equation}
\dif s^2=e^{2\kappa(\pi r_c-\abs{y})}\eta_{\mu\nu}\dif x^\mu\dif x^\nu+\dif y^2,
\end{equation}
where $y=r_c\theta,~\theta\in[-\pi,\pi],~r_c$ is the radius of the extra dimension, $\{x^\mu\},~\mu=0,1,2,3,$ are the coordinates in four-dimensional space-time, $\kappa$ is the scalar curvature in five dimensions, and $\eta_{\mu\nu}$ is the Minkowski metric. Thus, the space-time of the RS model is a slice of five-dimensional anti-de Sitter space ($\text{AdS}_5$) with a single compactified extra dimension. The periodicity condition for this extra dimension is $(x^\mu,y)=(x^\mu,y+2\pi r_c)$ and the points $(x^\mu,y)$ and $(x^\mu,-y)$ are equivalent. The main idea that underlies such form of the metric is to generate large hierarchy by adding an exponential function of the compactification radius to the metric. This means that one can attain a large ratio between the Planck and weak scales without requiring very large value for $r_c$.

In the following set-up we consider two three-dimensional branes with equal and opposite tensions---the first one (Plank brane) is situated at the point $y=0$ and the second one (TeV brane or visible brane) is located at $y=\pi r_c$. We assume that all SM fields are confined to the TeV brane whereas the gravity permeates all five dimensions. From the action of the effective four-dimensional theory one can derive a relation between the reduced four-dimensional Planck mass $\overbar{M}_\text{Pl}$ and the reduced gravity scale in five dimensions $\overbar{M}_5$ \cite{Randall:1999ee}:
\begin{equation}\label{eq:MPl-M5-rel}
\overbar{M}_\text{Pl}^2=\frac{\overbar{M}_5^3}{\kappa}\left(e^{2\pi\kappa r_c}-1\right).
\end{equation}

For an observer living on the TeV brane, the masses of the Kaluza-Klein (KK) graviton excitations depend linearly on the scalar curvature $\kappa$:
\begin{equation}
m_n=x_n\kappa,\quad n=1,2,\ldots,
\end{equation}
where $x_n$ are zeros of the Bessel function of the first kind $J_1(x)$. The coupling of the massless KK graviton $G_{\mu\nu}^{(0)}$ and massive KK gravitons $G_{\mu\nu}^{(n)}$, and the energy-momentum tensor $T^{\mu\nu}$ on the TeV brane, is given by interaction Lagrangian
\begin{equation}
\mathcal{L}_\text{int}=-\frac{1}{\overbar{M}_\text{Pl}}T^{\mu\nu}G_{\mu\nu}^{(0)}-\frac{1}{\Lambda_\pi}T^{\mu\nu}\sum_{n=1}^\infty G_{\mu\nu}^{(n)},
\end{equation}
where $\Lambda_\pi=(\overbar{M}_5^3/\kappa)^{1/2}$ sets a physical scale on the visible brane. In contrast to the standard RS model \cite{Randall:1999ee} in which the lightest KK excitation has mass of the order of 1 TeV, one can consider a scenario with small curvature, i.e. $\kappa\sim 1~\text{GeV},~\overbar{M}_5\sim 1~\text{TeV}$. As a result, the model contains infinite spectrum of closely spaced low-mass KK gravitons whose lightest mass is equal to $3.83\kappa$.

In order to be accepted or rejected, the RS model with a large extra dimension has to be experimentally tested. A few collaborations have examined the conceivable presence of compact extra dimension with the result that they determined lower bounds for the reduced gravity scale in five dimensions $\overbar{M}_5$. By studying the photon energy spectrum (respectively the missing transverse energy $\cancel{E}_{_{\!\perp\!}}$) in the process $e^+e^-\rightarrow\gamma+\cancel{E}_{_{\!\perp\!}}$ the DELPHI collaboration \cite{Ask:2007ia} obtained a limit on $\overbar{M}_5$: $\overbar{M}_5>0.92~\text{TeV}$ at 95\% CL. Furthermore, the search for large extra dimensions in the diphoton channel is carried out by the D\cancel{O} \cite{Abbott:2000zb} and CDF \cite{Acosta:2004sn} collaborations at $\sqrt{s}=1.96~\text{TeV}$. The bound that is obtained using their data is: $\overbar{M}_5>0.81~\text{TeV}$. Finally, studying the diphoton production at the LHC \cite{Kisselev:2008xv}, the discovery limits on $\overbar{M}_5$ are estimated to be 5.1 TeV and 6.3 TeV for integrated luminosities $30~\text{fb}^{-1}$ and $100~\text{fb}^{-1}$ respectively.

\section{Neutrino-nucleon inelastic cross-section in the RS model with a small curvature}

It is feasible to detect effects induced by low-mass KK gravitons by searching for their imprints in the scattering of the SM fields. To do that, we have to look at the trans-Planckian kinematical region
\begin{equation}\label{eq:kinreg}
\sqrt{s}\gtrsim\overbar{M}_5\gg-t,
\end{equation}
where $\sqrt{s}$ is center-of-mass energy and $t=-q_{_{\!\perp\!}}^2$ is four-dimensional momentum transfer. We also assume the small curvature option of the RS model, i.e. $\kappa\ll\overbar{M}_5$. Note that our assumptions do not contradict in any way the relation \eqref{eq:MPl-M5-rel}. According to the eikonal approximation, which is applicable in the regime \eqref{eq:kinreg}, elastic scattering amplitude is given by the sum of reggeized gravitons (gravi-Reggeons) in the $t$-channel. The presence of extra dimension leads to splitting of the Regge trajectory into an infinite sequence of trajectories enumerated by the KK number $n$ \cite{Kisselev:2008jw}:
\begin{equation}
\alpha_n(t)=2+\alpha_g^\prime t-\alpha_g^\prime m_n^2,\quad n=0,1,\ldots
\end{equation}
According to the string theory, the slope of the gravi-Reggeons is universal for all $s$ and is given by the string scale $M_\text{s}$ via $\alpha_g^\prime=1/M_\text{s}^2$. For numerical calculations we choose $M_\text{s}=1~\text{TeV}$.

The differential neutrino-nucleon cross section is derived in eikonal approximation \cite{Kisselev:2008jw,Kisselev:2005yn}:
\begin{equation}
\label{eq:crosssection}
\od{\sigma}{y}=\frac{1}{16\pi s}\abs{A_{\nu N}(s,t)}^2,
\end{equation}
where the neutrino-nucleon amplitude $A_{\nu N}$ is expressed in terms of the eikonal $\chi$:
\begin{equation}
A_{\nu N}(s,t)=4\pi i\,s\int\limits_0^\infty\dif b\,bJ_0(bq_{_{\!\perp\!}})\left\{1-\exp\left[i\chi(s,b)\right]\right\}.
\end{equation}
The eikonal is defined by the hadronic Born amplitude
\begin{equation}
\chi(s,b)=\frac{1}{4\pi s}\int\limits_0^\infty\dif q_{_{\!\perp\!}}\,q_{_{\!\perp\!}}J_0(q_{_{\!\perp\!}}b)A_{\nu N}^\text{B}(s,t)
\end{equation}
which, in its turn, is given by the gravity Born amplitude for neutrino scattering off a point-like particle and the $t$-dependent parton distributions:
\begin{equation}
A_{\nu N}^\text{B}(s,t)=\sum_{i=q,\overbar{q},g}\int\limits_0^1\dif x\,A_\text{grav}^\text{B}(xs,t)F_i(x,t),
\end{equation}
\begin{equation}
A_\text{grav}^\text{B}(s,t)=\frac{\pi\alpha_g^\prime s^2}{2\Lambda_\pi^2}\sum_{n\neq 0}\left[i-\cot\frac{\pi\alpha_n(t)}{2}\right]\left(\frac{s}{\overbar{M}_5}\right)^{\alpha_n(t)-2}.
\end{equation}
The parton distributions have a Regge-like form
\begin{equation}
\label{eq:partondist}
F_i(x,t)=f_i(x)\exp\left[t(r_0^2-\alpha_\text{P}^\prime\ln x)\right]
\end{equation}
with parameters $r_0^2=0.62~\text{GeV}^{-2}$ and $\alpha_\text{P}^\prime=0.094~\text{GeV}^{-2}$ \cite{Petrov:2001eu}. For our calculations we use a set of parton distribution functions $f_i(x)$ from \cite{cteq}.

\begin{figure}[h!]
\centering
\includegraphics[width=0.8\textwidth]{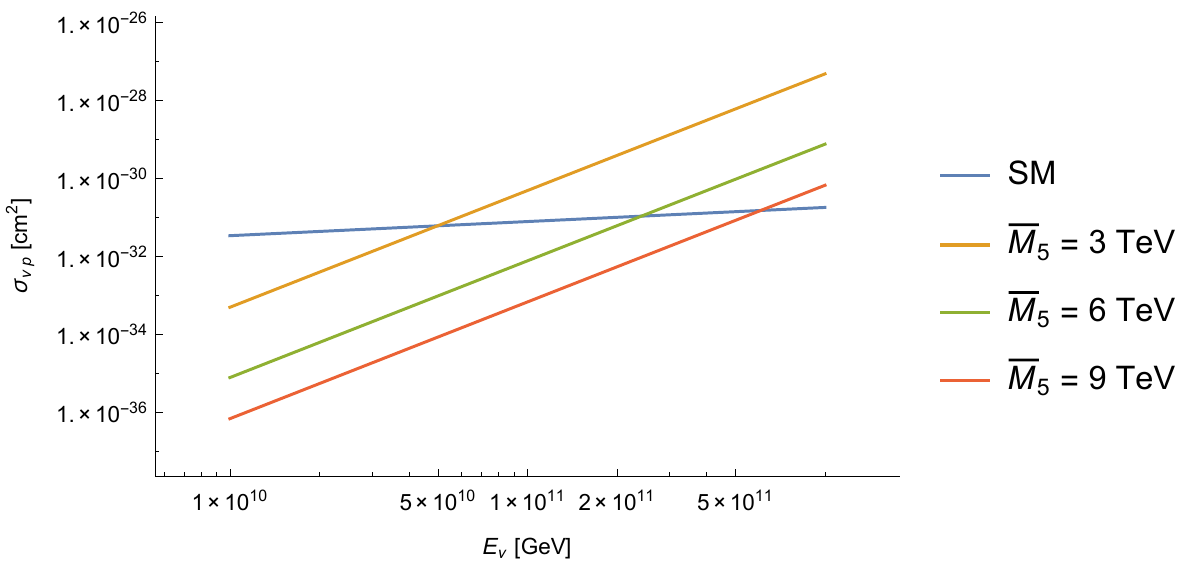}
\caption{Total neutrino-nucleon cross sections as a function of neutrino energy for $\kappa=1~\text{GeV}$ and three different values of $\overbar{M}_5$. For comparison: SM extrapolation.}
\label{fig:crosssectionsplot}
\end{figure}

On Fig. \ref{fig:crosssectionsplot} we plot the total neutrino-nucleon cross section calculated by making use of Eqs. \eqref{eq:crosssection}--\eqref{eq:partondist} for $\kappa=1~\text{GeV}$ and three different values of the reduced gravity scale $\overbar{M}_5$. For comparison is plotted the SM extrapolation taken from \cite{Gandhi:1998ri} and valid in the region $10^7~\text{GeV}\leq E_\nu\leq 10^{12}~\text{GeV}$ within 10\%:
\begin{equation}
\sigma_{\nu N}^\text{tot}=7.84\times 10^{-36}~\text{cm}^2\left(\frac{E_\nu}{1~\text{GeV}}\right)^{0.363}.
\end{equation}
One can readily notice that the total neutrino-nucleon cross section predicted by the RS model sufficiently dominates the SM one at high energies. This fact indicates that a large-exposure cosmic-based detector with high energy threshold can be sensitive to gravity effects in deeply penetrating quasi-horizontal air showers induced by ultra-high energy cosmic neutrinos.

\section{Detection of quasi-horizontal air showers with JEM-EUSO}

The neutrinos, in particular, are characterized by specific pattern of the air showers they induce in the Earth atmosphere compared to the protons for example. This is due to the fact that neutrinos have extremely small cross section for interaction with atmospheric nuclei. For this reason neutrinos must pass much more atmosphere in order to interact and initiate air showers. Thus quasi-horizontal air showers are needed to detect high-energy neutrinos.

The Extreme Universe Space Observatory on-board the Japanese Experiment Module (JEM-EUSO) \cite{Takahashi:2009,Ebisuzaki:2014} on the International Space Station (ISS) is an innovative space mission designed to detect ultra-high energy cosmic rays (UHECRs). Orbiting the Earth with period of 90 minutes, it uses large volume of the Earth atmosphere as a detector. When a high-energy particle enters the atmosphere, it interacts with the atmospheric nuclei thus initiating Extensive Air Shower (EAS) of secondary charged particles. The JEM-EUSO telescope exploits the isotropic fluorescence light emitted from the EAS as well as the Cherenkov light reflected from Earth surface or thick clouds. Using that technique, JEM-EUSO works as ultrafast camera that records the spatial and temporal profile of the EAS during its development in the atmosphere. The main objective of the mission is UHECR astronomy and astrophysics.

In order to calculate the expected number of neutrino-induced air showers that the JEM-EUSO telescope can possibly detect, we need to know the cosmic neutrino flux $\Phi_\nu(E_\nu)$ and the annual exposure $\Psi(E_\nu)$ of the telescope. For the cosmic neutrino flux we take the optimistic Waxman-Bahcall upper bound $E_\nu^2\Phi_\nu<2\times 10^{-8}~\text{GeV}\,\text{cm}^{-2}\,\text{s}^{-1}\,\text{sr}^{-1}$ \cite{Waxman:1998yy,Bahcall:1999yr}.

\begin{figure}[h!]
\centering
\includegraphics[width=0.8\textwidth]{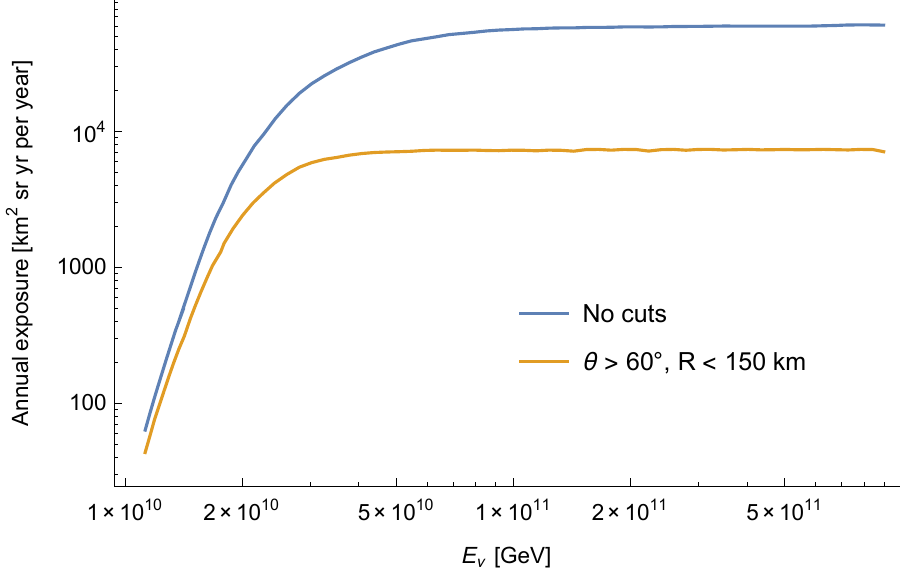}
\caption{Annual exposure of the JEM-EUSO observatory as function of neutrino energy $E_\nu$ for ISS altitude of 400 km (full FoV and $\theta>60\degree,~R<150~\text{km}$ cut) \cite{Adams:2013vea}.}
\label{fig:exposureplot}
\end{figure}

The nominal instantaneous aperture of JEM-EUSO in nadir mode is $\sim 5\times 10^5~\text{km}^2\,\text{sr}$ at the highest energies. A detailed study of all factors that affect the annual exposure has been conducted by the JEM-EUSO collaboration \cite{Adams:2013vea} (see Fig. \ref{fig:exposureplot}). These factors give conversion coefficient between geometrical aperture and exposure of about $\sim 13\%$.

For numerical calculations we take the threshold energy of the JEM-EUSO detector to be $E_\text{th}=3\times10^{10}~\text{GeV}$ and integrate up to maximal energy $E_\text{max}=1\times10^{12}~\text{GeV}$. The number of neutrino-induced quasi-horizontal air showers per year is given by \cite{PalomaresRuiz:2005xw}

\begin{equation}
N_{e\nu}~[\text{yr}^{-1}]=hN_\text{A}\rho_\text{atm}(0)\int\limits_{E_\text{th}}^{E_\text{max}}\dif E_\nu\,\Phi_\nu(E_\nu)\Psi(E_\nu)\sigma_{\nu N}(E),
\end{equation}
where $h$ is the atmospheric scale height, $N_\text{A}$ is the Avogadro constant, and $\rho_\text{atm}(0)$ is the air density at sea level. Since the neutrinos are characterized by quasi-horizontal air showers, it is reasonable to apply simple geometrical cuts on the annual exposure of JEM-EUSO. The particular cut we will use is on the zenith angle $\theta>60\degree$ and on the distance $R<150~\text{km}$ from the center of FoV of the impact location of the EAS (orange line in Fig. \ref{fig:exposureplot}).

\begin{figure}[h!]
\centering
\includegraphics[width=0.75\textwidth]{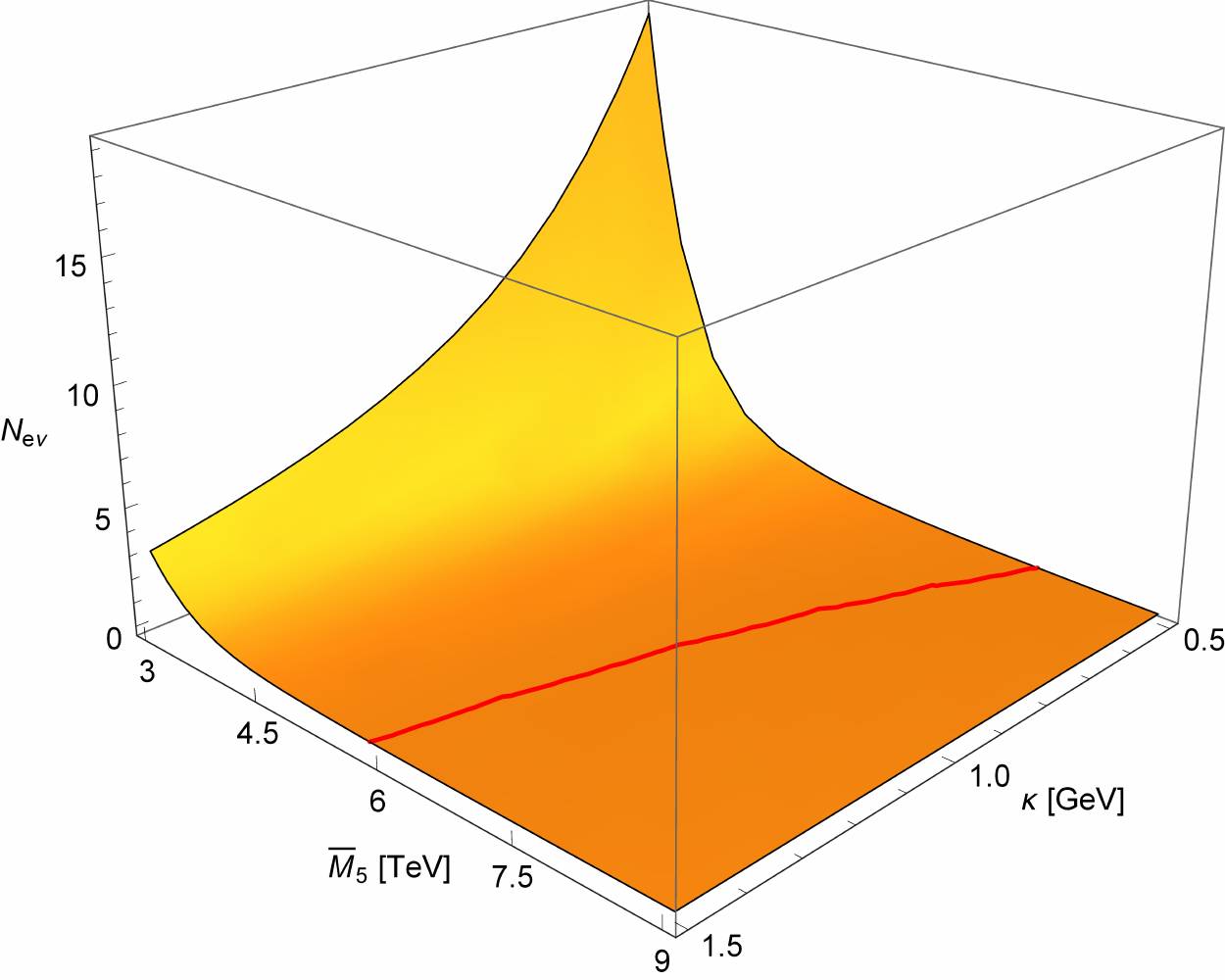}
\caption{Expected number of neutrino-induced quasi-horizontal air showers detected for a year by the JEM-EUSO observatory as a function of the parameters $\kappa$ and $\overbar{M}_5$. The thick red curve: SM expectation.}
\label{fig:showersplot}
\end{figure}

The results of our calculations are shown in Table \ref{tab:numshowers} and in Fig. \ref{fig:showersplot} for different values of two parameters of the model.

\begin{table}[h!]
\centering
\begin{tabular}{ c c c c c }
\toprule
& $\overbar{M}_5=3~\text{TeV}$ & $\overbar{M}_5=4~\text{TeV}$ & $\overbar{M}_5=5~\text{TeV}$ & SM\\
\midrule
$\kappa=0.5~\text{GeV}$ & 19.14 & 3.41 & 0.89 \\
$\kappa=1.0~\text{GeV}$ & 6.72 & 1.20 & 0.31 & 0.06 \\
$\kappa=1.5~\text{GeV}$ & 3.21 & 0.57 & 0.15 \\
\bottomrule
\end{tabular}
\caption{Expected number of neutrino-induced quasi-horizontal air showers at the JEM-EUSO observatory for several values of the parameters $\kappa$ and $\overbar{M}_5$.}
\label{tab:numshowers}
\end{table}

\section{Conclusion}

The event rate of neutrino-induced quasi-horizontal air showers depends on two parameters of the Randall-Sundrum model---the scalar curvature $\kappa$ and the reduced gravity scale $\overbar{M}_5$ in five dimensions. Consequently, there exists a sufficient difference in the event rates as predicted by the RS model and by the SM (Table \ref{tab:numshowers} and Fig. \ref{fig:showersplot}). Such deviation from the SM expectation can be searched out by making use of a large-aperture detector that is able to detect the weakly interacting high-energy cosmic neutrinos. For this reason we have calculated the expected number of neutrino-induced quasi-horizontal EAS that the forthcoming JEM-EUSO mission is capable of detecting for a year. Taking into account that JEM-EUSO can reach an annual exposure one order of magnitude higher than the Pierre Auger Observatory, it will be possible to set more accurate lower bounds on the parameters of the RS model.

\section{Acknowledgments}
We express our gratitude to the Research fund of Sofia University for the support of this work through Grant Agreement FNI-SU 53/3.4.15. This work was supported under the ESA Topical Team Contract No. 4000103396 and by the Bundesministerium für Wirtschaft und Technologie through the Deutsches Zentrum für Luft- und Raumfahrt e.V. (DLR) under the grant number FKZ 50 QT 1101.


\begin{thebibliography}{99}

\bibitem{ArkaniHamed:1998rs} 
  N.~Arkani-Hamed, S.~Dimopoulos and G.~R.~Dvali,
  Phys.\ Lett.\ B {\bf 429}, 263 (1998)
  \href{http://arxiv.org/abs/hep-ph/9803315}{[hep-ph/9803315]}.

\bibitem{Antoniadis:1998ig} 
  I.~Antoniadis, N.~Arkani-Hamed, S.~Dimopoulos and G.~R.~Dvali,
  Phys.\ Lett.\ B {\bf 436}, 257 (1998)
  \href{http://arxiv.org/abs/hep-ph/9804398}{[hep-ph/9804398]}.

\bibitem{ArkaniHamed:1998nn} 
  N.~Arkani-Hamed, S.~Dimopoulos and G.~R.~Dvali,
  Phys.\ Rev.\ D {\bf 59}, 086004 (1999)
  \href{http://arxiv.org/abs/hep-ph/9807344}{[hep-ph/9807344]}.

\bibitem{Randall:1999ee} 
  L.~Randall and R.~Sundrum,
  Phys.\ Rev.\ Lett.\  {\bf 83}, 3370 (1999)
  \href{http://arxiv.org/abs/hep-ph/9905221}{[hep-ph/9905221]}.

\bibitem{Ask:2007ia} 
  S.~Ask {\it et al.}  [DELPHI Collaboration],
  \href{http://arxiv.org/abs/0707.2102}{arXiv:0707.2102 [hep-ex]}.

\bibitem{Abbott:2000zb} 
  B.~Abbott {\it et al.}  [D0 Collaboration],
  Phys.\ Rev.\ Lett.\  {\bf 86}, 1156 (2001)
  \href{http://arxiv.org/abs/hep-ex/0008065}{[hep-ex/0008065]}.

\bibitem{Acosta:2004sn} 
  D.~Acosta {\it et al.}  [CDF Collaboration],
  Phys.\ Rev.\ Lett.\  {\bf 95}, 022003 (2005)
  \href{http://arxiv.org/abs/hep-ex/0412050}{[hep-ex/0412050]}.

\bibitem{Kisselev:2008xv} 
  A.~V.~Kisselev,
  JHEP {\bf 0809}, 039 (2008)
  \href{http://arxiv.org/abs/0804.3941}{[arXiv:0804.3941 [hep-ph]]}.

\bibitem{Kisselev:2008jw} 
  A.~V.~Kisselev,
  Open Astron.\ J.\  {\bf 2}, 12 (2009)
  \href{http://arxiv.org/abs/0807.3307}{[arXiv:0807.3307 [hep-ph]]}.

\bibitem{Kisselev:2005yn} 
  A.~V.~Kisselev and V.~A.~Petrov,
  Phys.\ Rev.\ D {\bf 71}, 124032 (2005)
  \href{http://arxiv.org/abs/hep-ph/0504203}{[hep-ph/0504203]}.

\bibitem{Petrov:2001eu} 
  V.~A.~Petrov and A.~V.~Prokudin,
  Eur.\ Phys.\ J.\ C {\bf 23}, 135 (2002)
  \href{http://arxiv.org/abs/hep-ph/0105209}{[hep-ph/0105209]}.

\bibitem{cteq}
  \url{http://users.phys.psu.edu/~cteq/#PDFs}

\bibitem{Gandhi:1998ri} 
  R.~Gandhi, C.~Quigg, M.~H.~Reno and I.~Sarcevic,
  Phys.\ Rev.\ D {\bf 58}, 093009 (1998)
  \href{http://arxiv.org/abs/hep-ph/9807264}{[hep-ph/9807264]}.

\bibitem{Takahashi:2009}
  Y.~Takahashi {\it et al.} (JEM-EUSO Coll.),
  New Journal of Physics {\bf 11}, 065009 (2009).

\bibitem{Ebisuzaki:2014}
  T.~Ebisuzaki {\it et al.} (JEM-EUSO Coll.),
  Advances in Space Research, Vol. {\bf 53}, Issue 10, Pages 1499--1505 (2014).

\bibitem{Waxman:1998yy} 
  E.~Waxman and J.~N.~Bahcall,
  Phys.\ Rev.\ D {\bf 59}, 023002 (1999)
  \href{http://arxiv.org/abs/hep-ph/9807282}{[hep-ph/9807282]}.

\bibitem{Bahcall:1999yr} 
  J.~N.~Bahcall and E.~Waxman,
  Phys.\ Rev.\ D {\bf 64}, 023002 (2001)
  \href{http://arxiv.org/abs/hep-ph/9902383}{[hep-ph/9902383]}.


\bibitem{Adams:2013vea} 
  J.~H.~Adams {\it et al.}  [JEM-EUSO Collaboration],
  Astropart.\ Phys.\  {\bf 44}, 76 (2013)
  \href{http://arxiv.org/abs/1305.2478v1}{[arXiv:1305.2478 [astro-ph.HE]]}.

\bibitem{PalomaresRuiz:2005xw} 
  S.~Palomares-Ruiz, A.~Irimia and T.~J.~Weiler,
  Phys.\ Rev.\ D {\bf 73}, 083003 (2006)
  \href{http://arxiv.org/abs/astro-ph/0512231v1}{[astro-ph/0512231]}.

\end{thebibliography}
\end{document}